\documentclass[american,reprint,aps,pra,superscriptaddress,showpacs]{revtex4-1}
\usepackage{amsthm}
\usepackage{amsmath}
\usepackage{amssymb}
\usepackage{graphicx, graphics,epstopdf}
\usepackage[unicode=true,pdfusetitle,
 bookmarks=true,bookmarksnumbered=false,bookmarksopen=false,
 breaklinks=false,pdfborder={0 0 0},backref=false,colorlinks=false]
 {hyperref}
\hypersetup{colorlinks,linkcolor=myurlcolor,citecolor=myurlcolor,urlcolor=myurlcolor}

\usepackage{times}
\usepackage{colortbl}\definecolor{myurlcolor}{rgb}{0,0,0.7}

\newcommand{\ket}[1]{|{#1}\rangle}
\newcommand{\bra}[1]{\langle{#1}|}
\newcommand{\braket}[2]{\langle {#1} | {#2} \rangle}
\newcommand{\ketbra}[2]{|{#1} \rangle \langle {#2} |}

\begin{document}
\title{Measuring Non-Hermitian Operators via Weak Values}
\author{Arun Kumar Pati}
\email{akpati@hri.res.in}

\author{Uttam Singh}
\email{uttamsingh@hri.res.in}

\affiliation{Harish-Chandra Research Institute, Allahabad 211 019, India}

\author{Urbasi Sinha}
\email{usinha@rri.res.in}

\affiliation{Raman Research Institute, Bangalore 560 080, India}


\begin{abstract}
In quantum theory, a physical observable is represented by a Hermitian operator as it admits real eigenvalues. This stems from the fact that any measuring apparatus that is supposed to measure a physical observable will always yield a real number. However, reality of eigenvalue of some operator does not mean that it is necessarily Hermitian. There are examples of non-Hermitian operators which may admit real eigenvalues under some symmetry conditions. However, in general, given a non-Hermitian operator, its average value in a quantum state is a complex number and there
are only very limited methods available to measure it. Following standard quantum mechanics, we provide an experimentally feasible protocol to measure the expectation value of any non-Hermitian operator via weak measurements. The average of a non-Hermitian operator in a pure state is a complex multiple of the weak value of the positive semi-definite part of the non-Hermitian operator. We also prove a new uncertainty relation for any two non-Hermitian operators and show that the fidelity of a quantum state under quantum channel can be measured using the average of the corresponding Kraus operators. The importance of our method is shown in testing the stronger uncertainty relation, verifying the Ramanujan formula and in measuring the product of non commuting projectors.
\end{abstract}

\maketitle

\section{Introduction}
One of the basic postulates of quantum mechanics limits the possible quantum mechanical observables to be the Hermitian ones \cite{Dirac1981}. The Hermiticity of the quantum mechanical observables seems to be a compelling and plausible postulate as the eigenvalues of Hermitian operators are real. Moreover, a Hermitian Hamiltonian yields a unitary evolution leading to the conservation of probability. But the reality of the spectrum of the quantum mechanical observables does not imply that the observables must be Hermitian. In fact, there are certain class of operators that are not Hermitian yet their spectrum is real. The reason for the reality of such operators is argued to be the underlying symmetry of the operators with certain other restrictions. This has resulted in the attempts to lift the postulate of Hermiticity and allow for more general operators. It is known that there are non-Hermitian operators which possess real eigenvalues if one imposes
some symmetry conditions, namely the $\mathcal{P}T-$symmetry, which is unbroken. $\mathcal{P}T-$symmetry is said to be not spontaneously broken if the eigenfunctions of the non-Hermitian operator are itself $\mathcal{P}T-$symmetric. Such kind of operators which respect the unbroken $\mathcal{P}T-$symmetry are ingredients of $\mathcal{P}T-$symmetric quantum mechanics \cite{Bender1998, Bender2002, Brody2012}.

In quantum theory, the concept of weak measurement was introduced by Aharonov-Albert-Vaidman \cite{Aharonov1998, Aharonov2005, Aharonov2008} to study the properties of a quantum system in pre and postselected ensembles. In this formalism the measurement of an observable leads to a weak value of the observable with unexpectedly strange properties. In fact, the weak value is shown to be complex, in general, and can take values outside the eigen-spectrum of the observable. The concept of weak measurements has been generalized further beyond its original formulation \cite{Dressel2010, Dressel2012, DresselP2012, Wu2011, Lorenzo2012, Pang2012, Kofman2012, Shikano2010, Shomroni2013}. In recent years, weak values have found numerous applications. For example, the Panchratnam geometric phase is nothing but the phase of a complex valued weak value that arises in the context of weak measurements \cite{Sjoqvist2006}. It has been shown that weak measurements can be used for interrogating quantum systems in a coherent manner \cite{Aharonov2010, Cho2011}. In addition, it plays important role in understanding the uncertainty principle in the double-slit experiment \cite{Wiseman2003, Mir2007}, resolving Hardy's paradox \cite{Lundeen2009}, analyzing tunneling time \cite{Steinberg1995, SteinbergP1995}, protecting quantum entanglement from decoherence \cite{Kim2012, Singh2014}, modifying the decay law \cite{Davies2009}. Remarkably, it is possible to express the wavefunction as a weak value of a projector and this paved the way to measure the wavefunction of single photon directly \cite{Lundeen2011,Lundeen2012}.  Similarly, in quantum metrology the phase sensitivity of a quantum measurement is given by the variance of the imaginary parts of the weak values of the generators over the different measurement outcomes \cite{Hofmann2011}. For a very recent review on weak measurements one can look at Ref. \cite{Dressel2014}. 

Despite having complex spectrum, in general, the non-Hermitian operators have found applications in theoretical work as a mathematical model for studying open quantum systems in nuclear physics \cite{Feshbach1958} and quantum optics \cite{Plenio1998}, among others to name a few. In these fields, the non-Hermitian Hamiltonian appears as an effective description for the subsystem of the full system. The adiabatic measurements on systems evolving according to effective non-Hermitian Hamiltonians are analyzed in Ref. \cite{Aharonov1996} and it is established that the outcome of an adiabatic measurement of a Hermitian observable is the weak value associated with the two-state vector comprising of forward and backward evolving eigenstates of the non-Hermitian Hamiltonian. Non-Hermitian operators that can be expressed as a product of two non commuting Hermitian operators do appear in the formalism that describes quantum states using quasiprobability distribution such as the Dirac distribution \cite{Dirac1945, Chaturvedi2006, Johansen2007, Bamber2014}, the Moyal distribution \cite{Moyal1949, DiLorenzo2013}, etc. Also, weak measurement of a Hermitian operator on a system having non-Hermitian Hamiltonian is considered in Ref. \cite{Matzkin2012}. Apart from these limited expositions of the measurement of non-Hermitian operators, not much is discussed towards experimental methods to measure the expectation values of such operators.
%
%
Here we propose an experimentally verifiable procedure to measure the complex expectation value of a general non-Hermitian operator. The key to measurement of non-Hermitian operators is the notion of the polar decomposition of any operator and the process of weak measurement. In this paper, we show that
the average of a non-Hermitian operator  is a complex multiple of the weak value of the positive
semi-definite part of the non-Hermitian operator. 
By experimentally verifiable procedure to measure non-Hermitian operators
we mean that the expectation value of a non-Hermitian operator in a
quantum state is inferred from a direct measurement of the weak value of
its Hermitian positive definite part employing the theoretically
determined expectation value of the  unitary part of the non-Hermitian
operator.
Significantly, our method can be used to measure the matrix elements of any non-Hermitian operator.
It is important to note that our proposed method to measure the expectation value require a priori knowledge of both the operator to be measured and the state in which it is being measured. However, this situation may arise naturally in many contexts, therefore it does not limit the applicability of our method severely. Besides the mentioned situations in quantum optics where non-Hermitian operators appear, one noteworthy example is the quantum system described by evolution in presence of gain or loss \cite{Brody2012}.
Also, we would like to mention here that our approach is different from the one that is used in the direct measurement of wavefunction \cite{Lundeen2011, Lundeen2012} and goes beyond it as the former is applicable to any non-Hermitian operator.
Moreover, we prove a new uncertainty relation for any two non-Hermitian operators and show that it can also be tested experimentally. As an application, we show that the uncertainty in the Kraus operators governs the fidelity of the output state for a quantum channel. If the total uncertainties in the Kraus operators is less, then the fidelity will be more. We illustrate our main results with several examples. Our method allows one to measure the average of creation and annihilation operators in any state and provides an experimental method to test the Ramanujan formula for the sum of square roots of first $s$ natural numbers.

The paper is organized as follows: In Sec. \ref{sec:NH} we give a very brief review of non-Hermitian operators in quantum mechanics. We discuss our method to measure expectation values of non-Hermitian operators in Sec. \ref{exp-non}. In Sec. \ref{uncer-non}, we find an uncertainty relation for non-Hermitian operators and using the results of the previous section, we make a connection of the uncertainty relation to experiments. To exemplify this, we provide the uncertainty relation for creation and annihilation operators.  In Sec. \ref{uncer-kraus}, we find a relation between fidelity of a quantum channel and uncertainty of non-Hemitian Kraus operators of the channel and provide an example of amplitude damping channel as an illustration.  In Sec. \ref{apps}, we give a comprehensive account of the interesting applications of our results in the context of testing stronger uncertainty relation, measuring product of projection operators and in verifying Ramanujan's formula. Finally, we conclude in Sec. \ref{conclusion} together with the discussion and implications of our results.

\section{Non-Hermitian operators}
\label{sec:NH}
For the sake of clarity and completeness, in this section we will review the non-Hermitian operators in quantum mechanics. The abstract mathematical description of quantum mechanics is facilitated by the introduction of a separable Hilbert space $\mathcal{H}$ which by definition is complete and endowed with an inner product \cite{Merzbacher1998}. The states of the physical system and the physical observables are mapped one to one to the rays in the Hilbert space and Hermitian operators on the Hilbert space, respectively. The physically measurable quantity associated to the operator $O$ of a system in state $\ket{\psi} \in \mathcal{H}$ is the expectation value $\bra{\psi}O\ket{\psi}$. It is assumed here that the operator $O$ is Hermitian so that its expectation value is real. It is worth pointing out that the notion that an operator is Hermitian or not depends on an inner product, e.g. $O$ is Hermitian if $O=O^\dag$ where adjoint $O^\dag$ of $O$ is defined as
\begin{align}
 \bra{\psi}O\ket{\phi} = \bra{O^\dag \psi}{\phi}\rangle,
\end{align}
for all $\ket{\psi}\in \mathcal{H}$ and $\ket{\phi}\in \mathcal{H}$ in the domain of $O$. A closely related concept is of a self adjoint operator which is a Hermitian operator with the same domain for its adjoint \cite{Reed1980}. With a given Hilbert space and inner product defined over it, the operators $O$ for which $O\neq O^\dag$ is called as a non-Hermitian operator. Since the notion of Hermiticity or non-Hermiticity is relative to some inner product, a non-Hermitian operator relative to some inner product can be turned into a Hermitian operator relative to some other inner product. A very good exposition of this fact can be found in Refs. \cite{Mostafazadeh2010, Moiseyev2011, Croke2015}. The non-Hermiticity arising from the representation of an operator by square integrable functions such that these functions are not in the domain of the operator is not considered here.
%
%
In this work we focus on the measurability of non-Hermitian operators keeping the notion of inner product fixed once and for all. The non-Hermitian operators naturally occur in effective descriptions of open quantum systems \cite{Feshbach1958}, systems in presence of loss and gain \cite{Brody2012} and quantum optics \cite{Plenio1998}, etc (see also \cite{Moiseyev2011}). In general, the expectation value of a non-Hermitian operator in some state is a complex number. Various physical interpretations of the complex expectation of non-Hermitian operators in a quantum state are discussed in Ref. \cite{Moiseyev2011} for example in the scattering experiments. The adiabatic measurements with an experimental proposal on systems evolving according to effective non-Hermitian Hamiltonians can be found in Ref. \cite{Aharonov1996} and the results on weak measurement of a Hermitian operator on a system having non-Hermitian Hamiltonian can be found in Ref. \cite{Matzkin2012}.

In the next section, we provide our protocol to measure the complex expectation values of non-Hermitian operators employing weak measurements.

\section{ Expectation value of a non-Hermitian operator}
\label{exp-non}
Let us consider a non-Hermitian operator $A$. The expectation value of such an operator in a quantum state $\ket{\psi}$, given by $\bra{\psi} A\ket{\psi}$, is in general a complex number. This makes it unobservable in a laboratory experiment. But here we present a formalism to overcome this problem. To present the main idea, we need the polar decomposition of a matrix. Let $A \in \mathbb{ C}^{m \times n}, m \ge n$. Then, there exists a matrix $U \in  \mathbb{C}^{m \times n}$ and a unique Hermitian positive semi-definite matrix $R \in  \mathbb{C}^{n \times n}$ such that $A = UR$, with $U^{\dagger} U= I$. The positive semi-definite matrix $R$ is given by $R = \sqrt{A^{\dagger} A}$, even if $A$ is singular. If rank ($A$) = $n$ then $R$ is positive definite and $U$ is uniquely determined \cite{Higham1986, Gantmacher1959}.

Let us consider a quantum system initially in the state $\ket{\psi} \in {\cal H} = \mathbb{C}^d$. Suppose we are interested in measuring the average of a non-Hermitian operator $A$ in the state $\ket{\psi}$. Consider the polar decomposition of an operator $A \in  \mathbb{C}^{d\times d}$, given by $A=UR$, where $R$ is a positive semi-definite operator and $U$ is a unitary operator. The average of a non-Hermitian operator in the pure state $\ket{\psi}$, is given by 
\begin{align}
 \langle A\rangle &= \bra{\psi} A \ket{\psi} =  \bra{\psi} UR \ket{\psi}\nonumber\\
 &=\frac{\bra{\phi} R \ket{\psi}}{\langle \phi \ket{\psi}} \langle \phi \ket{\psi} ={_{\phi}\langle} R \rangle^w_{\psi} \langle \phi \ket{\psi},
\end{align}
where $\ket{\phi} = U^\dagger\ket{\psi}$ and ${_{\phi}\langle} R \rangle^w_{\psi}$ is the weak value of positive semi-definite operator $R$, given by $\frac{\bra{\phi} R \ket{\psi}}{\langle \phi \ket{\psi}}$. Now, given a non-Hermitian operator $A$, we first find out $R=\sqrt{A^\dagger A}$ and the corresponding unitary $U$. The measurement of the expectation value of $A$ in a quantum state $\ket{\psi}$ can be carried out as follows. We start with a quantum system which is preselected in the state $\ket{\psi_i} = \ket{\psi}$ and weakly measure the positive semi-definite operator 
$R$ in the preselected state $\ket{\psi}$. The weak measurement can be realized using the interaction between the system and the measurement apparatus which is governed by the interaction Hamiltonian
\begin{align}
 H_{int}= g \delta(t-t_0) R \otimes P,
\end{align}
where $g$ is the strength of the interaction that is sharply peaked at $t=t_0$, $R$ is an observable of the system and $P$ is that of the apparatus. Under the action of the interaction Hamiltonian, the system and apparatus evolve as
\begin{align}
 \ket{\psi} \otimes \ket{\Phi}  \rightarrow e^{-\frac{i}{\hbar}g R \otimes P} \ket{\psi} \otimes \ket{\Phi}.
\end{align}
Here $ \ket{\Phi}$ is the initial state of the apparatus.
After the weak interaction, we postselect the system in the state $\ket{\phi} = U^\dagger\ket{\psi}$ with the postselection probability given by $|\braket{\phi}{\psi}|^2(1+2g \mathrm{Im}{_{\phi}\langle} R \rangle^w_{\psi} \langle P \rangle)$. This yields the desired weak value of $R$, i.e., ${_{\phi}\langle} R \rangle^w_{\psi} =\frac{\bra{\phi} R \ket{\psi}}{\langle \phi \ket{\psi}}$.
Therefore, multiplying $\braket{\phi}{\psi}$ to ${_{\phi}\langle} R \rangle^w_{\psi}$, gives us $\braket{\psi}{A|\psi}$. 
To sum up, in order to measure the expectation value of a non-Hermitian operator $A$, with polar decomposition $A=UR$, in a state $\ket{\psi}$ we first determine experimentally the weak value of $R$ with pre and postselection in the states $\ket{\psi}$ and $U^\dagger\ket{\psi}$, respectively. Then multiplying the obtained weak value with the theoretically determined complex number, which is the expectation value of $U^\dagger$ in the state $\ket{\psi}$, gives the expectation value $\bra{\psi}A\ket{\psi}$ of the non-Hermitian operator $A$. To this end we have a provided a procedure to measure the expectation value of a non-Hermitian operator. 
Equivalently, one can also write $A = S U$, where $S = U R U^{\dagger} = \sqrt{AA^\dagger}$ is a positive semi-definite operator. In this case the average of $A$ in a pure state $\ket{\psi}$ is given by
\begin{align}
 \langle A\rangle &= \bra{\psi} A \ket{\psi} =  \bra{\psi} S U \ket{\psi}\nonumber\\
 &=\frac{\bra{\psi} S \ket{\chi}}{\langle \psi \ket{\chi}} \langle \psi \ket{\chi} ={_{\psi}\langle} S \rangle^w_{\chi} \langle \psi \ket{\chi},
\end{align}
where $\ket{\chi} = U\ket{\psi}$ and ${_{\psi}\langle} S \rangle^w_{\chi}$ is the weak value of positive semi-definite operator $S$, given by $\frac{\bra{\psi} S \ket{\chi}}{\langle \psi \ket{\chi}}$.
Following the same procedure as above one can measure the weak value of $S$ with preselection in the state $\ket{\chi} = U\ket{\psi}$ and postselection in the state $\ket{\psi}$. Now the weak value, ${_{\psi}\langle} S \rangle^w_{\chi}$, multiplied by $\langle \psi \ket{\chi}$ yields $\braket{\psi}{A|\psi}$. Furthermore, we have
\begin{align}
 \braket{\psi}{A|\psi} = {_{\psi}\langle} S \rangle^w_{\chi} \langle \psi \ket{\chi} ={_{\phi}\langle} R \rangle^w_{\psi} \langle \phi \ket{\psi}.
\end{align}
Interestingly, our method can also be applied to measure the weak value of any non-Hermitian operator $A$ in a preselected state $\ket{\psi}$ and postselected state $\ket{\psi'}$. Using the polar
decomposition of $A = UR$, the weak value of $A$ is given by
\begin{align}
\label{weaknh}
 {_{\psi'}\langle} A\rangle^w_{\psi} = \frac{\braket{\psi'}{UR|\psi}}{\braket{\psi'}{\psi}} = {_{\psi''}\langle} R\rangle^w_{\psi}.z,
\end{align}
where ${_{\psi''}\langle} R\rangle^w_{\psi} = \frac{\braket{\psi''}{R|\psi}}{\braket{\psi''}{\psi}}$ is the weak value of $R$ and $z = \frac{\braket{\psi''}{\psi}}{\braket{\psi'}{\psi}}$. Thus, the weak value of any non-Hermitian operator $A$ with the preselection in the state $\ket{\psi}$ and postselection in the states $\ket{\psi'}$ is equal to the weak value of $R$ with the preselection and the postselection in the states $\ket{\psi}$ and $\ket{\psi''}$, respectively, multiplied by the complex number $z$.

\section{ Uncertainty relation for non-Hermitian operators}
\label{uncer-non}
For any Hermitian operator, if we measure it in an arbitrary state, there will always be a finite uncertainty, unless the state is an eigenstate of the observable (Hermitian operator) that is being measured. Similarly, one can ask if there is an uncertainty associated to the measurement of any non-Hermitian operator. The variance of a non-Hermitian operator $A$ in a state $\ket{\psi}$ is defined as $\Delta A^2 := \langle \psi| (A^\dagger-\langle A^\dagger\rangle)(A-\langle A\rangle)| \psi \rangle$, where $\langle A\rangle = \bra{\psi}A\ket{\psi}$ and $\langle A^{\dagger} \rangle = \bra{\psi}A^{\dagger}\ket{\psi}$ \cite{Anandan1990}. Also, $\Delta A^2 = \bra{\psi} A^{\dagger} A \ket{\psi} -\bra{\psi} A^\dagger \ket{\psi} \bra{\psi} A \ket{\psi} = \langle f| f\rangle$,
where $\ket{f} = (A-\langle A\rangle)\ket{\psi}$. Even though $A$ is non-Hermitian, $\ket{f}$ is a valid quantum state as using the polar decomposition of $A = S_A U_A $ renders $\ket{f}$ as a linear combination of $\ket{\psi}$ and $S_A (U_A\ket{\psi})$. Similarly, we can define the uncertainty for the non-Hermitian operator $B$ as  $\Delta B^2 = \bra{\psi} B^{\dagger} B \ket{\psi} -\bra{\psi} B^\dagger \ket{\psi} \bra{\psi} B \ket{\psi}  = \langle g| g\rangle$, where $\ket{g} = (B-\langle B\rangle)\ket{\psi}$. Now we have
\begin{align}
\label{URR}
 \Delta A^2\Delta B^2 &= \langle f| f\rangle \langle g| g\rangle \geq |\langle f| g\rangle|^2,
\end{align}
whereby in the last line we have used the Cauchy-Schwarz inequality. Now let us use the polar decompositions of $A$ and $B$, namely, $A = S_A U_A$ and $B = S_B U_B$. Using these, we can simplify Eq. (\ref{URR}) and obtain
\begin{align}
 \Delta A \Delta B &\geq |\bra{\psi} U_A^\dagger S_A S_B U_B \ket{\psi}  - \bra{\psi} U_A^\dagger S_A \ket{\psi}\bra{\psi} S_B U_B \ket{\psi}|\nonumber\\
 &=|_{\phi}[S_A P  S_B]^w_{\chi}|\cdot|\bra{\phi}\chi\rangle|,
\end{align}
where $P = ( I- \ket{\psi}\bra{\psi})$,  $\ket{\phi} = U_A\ket{\psi}$ and $\ket{\chi} = U_B\ket{\psi}$. Thus, we have the generalized uncertainty relation for any two non-Hermitian operators as given by
\begin{align}
\label{rel}
 \Delta A \Delta B &\geq |_{\phi}[S_A P S_B]^w_{\chi}|\cdot|\bra{\phi}\chi\rangle|,
\end{align}
where $_{\phi}[S_A P  S_B]^w_{\chi}$ is the weak value of the non-Hermitian operator $S_A P S_B$ and it can be determined using our experimentally viable method. For the case of Hermitian operators,
we have the Robertson uncertainty relation \cite{Robertson1929}. 

As an example of the generalized uncertainty relation, given by Eq. (\ref{rel}), we consider the creation and annihilation operators, which are non-Hermitian, for a single mode electromagnetic field,
in the phase state $\ket{\theta_m}$ \cite{Pegg1988, Barnett2007}. The phase states are the eigenstates of the Hermitian phase operator. The phase operator arises in the context of the polar decomposition of creation and annihilation operators and it is known that the polar decomposition of creation and annihilation operators for the radiation field has difficulties related to the unitary
part of the decomposition \cite{Dirac1927, Louisell1963, Susskind1964}. This problem is addressed as the non-existence of Hermitian phase operator for the infinite dimensional Hilbert space  \cite{Nieto1968, Leblond1976}. The problem is resolved by taking Hilbert space to be finite dimensional and taking the limit at the end of all the calculations. In the finite dimensional Hilbert space there is a well defined Hermitian phase operator called as the Pegg-Barnett phase operator \cite{Pegg1988}.  This leads to the polar decompositions of the creation and the annihilation operators,
which are given by
\begin{align}
\label{polr-decop-cre-ann}
 \hat{a} &= e^{i\hat{\phi}_\theta} \sqrt{\hat{N}}~\mathrm{ and}\nonumber\\
 \hat{a}^\dagger &= \sqrt{\hat{N}} e^{-i\hat{\phi}_\theta},
 \end{align}
where $\hat{\phi}_\theta$ is the Hermitian phase operator,
\begin{align}
\label{phase-operator}
\hat{\phi}_\theta = \sum_{m = 0}^{s} \theta_m \ketbra{\theta_m}{\theta_m},
\end{align}
with $\theta_m = \theta_0 + 2m\pi/(s+1)$ and $\ket{\theta_m}$ are the orthonormal phase states, given by 
\begin{align}
\label{phase-states}
\ket{\theta_m}= (s+1)^{-1/2} \sum_{n = 0}^{s} e^{in\theta_m} \ket{n}.
\end{align}
The phase states satisfy $e^{\pm i\hat{\phi}_\theta} \ket{\theta_m} = e^{\pm i\theta_m} \ket{\theta_m}$. Now, the generalized uncertainty relation, given by Eq. (\ref{URR}), for the creation and annihilation operators  of a single mode electromagnetic field, in the phase state $\ket{\theta_m}$ reads as
\begin{align}
\label{uncera}
 \Delta \hat{a}^\dagger \Delta \hat{a} \geq |\langle (\hat{a}^\dagger)^2\rangle - \langle \hat{a}^\dagger \rangle^2|,
\end{align}
where $\Delta A = \sqrt {\braket{\theta_m}{ A^\dagger A |\theta_m} -\braket{\theta_m} { A^\dagger|\theta_m}\braket{\theta_m}{A|\theta_m} }$ with $A = a, a^\dagger$. Using the expressions $\braket{\theta_m}{ \hat{a}^\dagger \hat{a} |\theta_m} = \frac{s}{2}  = \braket{\theta_m}{ \hat{a} \hat{a}^\dagger |\theta_m}$, $ \braket{\theta_m}{ \hat{a} |\theta_m}^\dagger = \braket{\theta_m}{ \hat{a}^\dagger |\theta_m}  = (s+1)^{-1}e^{-i\theta_m}\sum_{n=0}^{s}\sqrt{n}$ and $ \braket{\theta_m}{ (\hat{a}^\dagger)^2 |\theta_m} = (s+1)^{-2}e^{-2i\theta_m}\sum_{n=0}^{s}\sqrt{n(n-1)} $, we have
\begin{align}
\label{lhs-uncer}
 \Delta \hat{a}^\dagger \Delta \hat{a} = (\Delta \hat{a}^\dagger)^2 = \left|\frac{s}{2} - (s+1)^{-2}\sum_{m,n=0}^{s}\sqrt{nm}\right|. 
\end{align}
and 
\begin{align}
\label{rhs-uncer}
 &|\langle (\hat{a}^\dagger)^2\rangle - \langle \hat{a}^\dagger \rangle^2| \nonumber\\
 &= (s+1)^{-1} \left| \sum_{n=0}^{s}\sqrt{n(n-1)} - (s+1)^{-1}\sum_{m,n=0}^{s}\sqrt{nm} \right|.
\end{align}

\begin{figure}
\centering
\includegraphics[width =65mm]{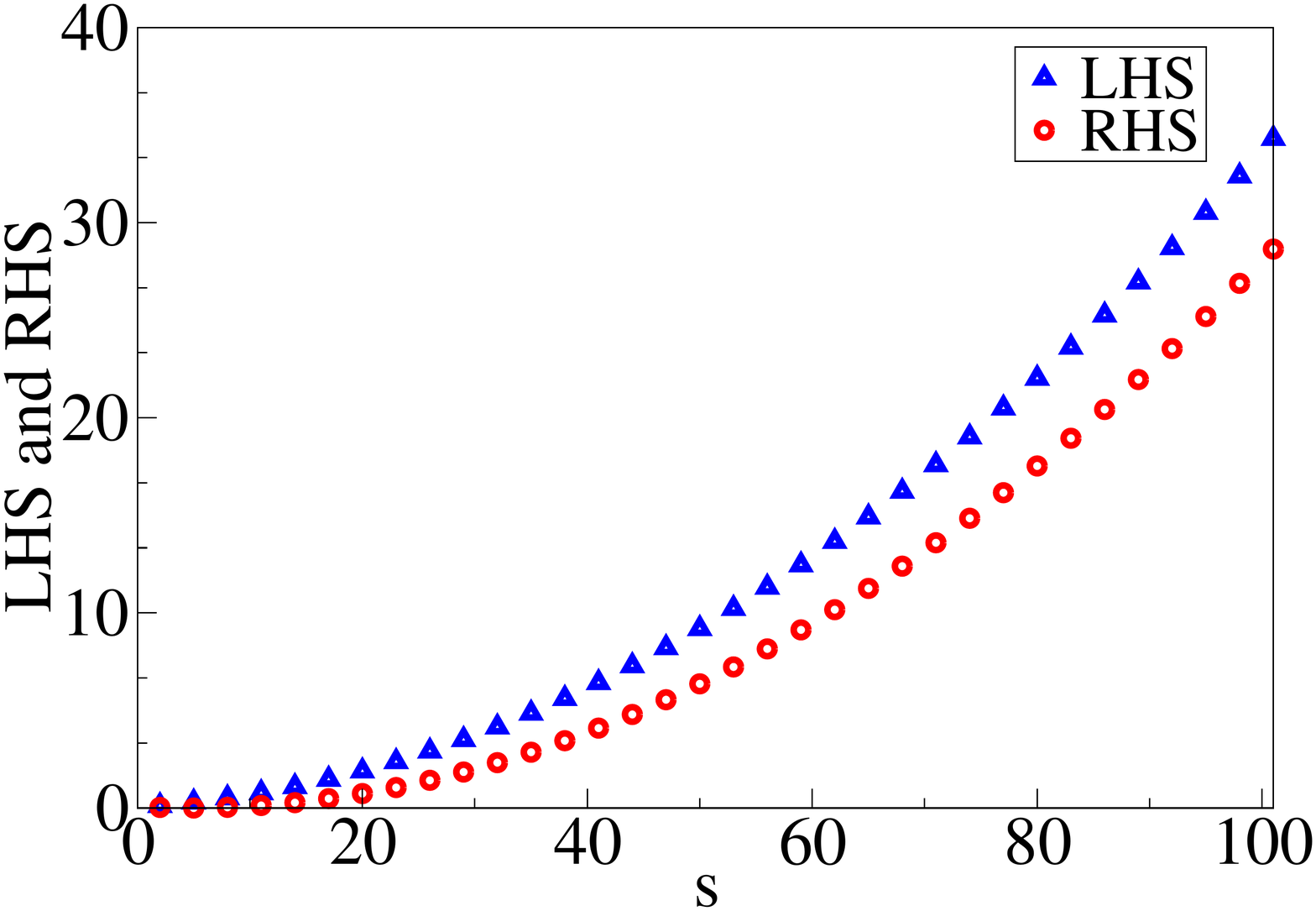}
\caption{(Color online) The plot of LHS and RHS of Eq. (\ref{uncera}), given respectively by Eq. (\ref{lhs-uncer}) and Eq. (\ref{rhs-uncer}), as a function of $s$. Both the axes are dimensionless. In the plot triangles show LHS and circles show RHS. This figure clearly shows that the uncertainty relation given for non-Hermitian operators is satisfied by the creation and annihilation operators in phase states.}
\label{fig1}
\end{figure}
One can see from Fig. (\ref{fig1}) that the uncertainty relation given by Eq. (\ref{uncera}) for creation and annihilation operators is indeed satisfied.

\section{Uncertainty for Kraus operators and fidelity of quantum states}
\label{uncer-kraus}
Here we will show how our proposal of measuring the average of non-Hermitian operator could be of high value. This will also show that the uncertainty in the non-Hermitian operator can have some real physical meaning. Suppose that a quantum system, initially in the state $\ket{\psi}$, passes through a quantum channel. The state of the system after passing through the quantum channel is given by
\begin{align}
\ketbra{\psi}{\psi} \rightarrow  \rho = {\cal E}(\ketbra{\psi}{\psi}) = \sum_{k} E_k \ketbra{\psi}{\psi}E_k^\dagger,
\end{align}
where $E_k's$ are the Kraus elements of the channel. Now, the fidelity between the pure initial state and the mixed final state is given by
\begin{align}
\label{fid}
 F = \bra{\psi} \rho \ket{\psi} = \sum_{k}|\braket{\psi}{E_k | \psi}|^2.
\end{align}
Eq. (\ref{fid}) shows that, by measuring the average of the non-Hermitian operators $E_k$ in the state $\ket{\psi}$, one can find the fidelity between the input and the output states.  Note that, usually, to measure the fidelity of a channel, one has to do a quantum state tomography of the final state and then calculate the quantity $\bra{\psi} \rho \ket{\psi}$. However, as stated above, by weakly measuring the positive semi-definite part of the Kraus operators, one can measure the average of the Karus operators and hence the channel fidelity.

Now consider the variance of $E_k$ in the state $\ket{\psi}$. This is given by
\begin{align}
 \Delta E_k^2 = \braket{\psi}{E_k^\dagger E_k|\psi } - \braket{\psi }{E_k^\dagger|\psi }\braket{\psi}{E_k|\psi }.
\end{align}
Summing both the sides and using the relation $\sum_k E_k^{\dagger} E_k = I$, we have 
\begin{align}
F +  \sum_k \Delta E_k^2  = 1.
\end{align}
This relation gives us a physical meaning to the uncertainties in the Kraus operators. This shows that if the total uncertainty in the Kraus operators is less, then the fidelity between the input and output
states will be more. Thus, the fidelity and the uncertainty play a complementary role in the quantum channel. Hence, to preserve a state more efficiently, one should have less uncertainties in the Kraus operators. For a quantum channel with two Kraus elements, the fidelity $F$ and uncertainties of the Kraus operators satisfy
\begin{align}
\label{final1}
  \frac{1- F }{2} \geq \Delta E_1\Delta E_2 \geq   |_{\phi}[S_1 \ket{\psi^\perp}\bra{\psi^\perp} S_2]^w_{\chi}|\cdot|\bra{\phi}\chi\rangle|,
\end{align}
where $E_1 = S_1U_1$, $E_2 = S_2U_2$, $\ket{\phi} = U_1\ket{\psi}$, $\ket{\chi} = U_2\ket{\psi}$, and $\ketbra{\psi}{\psi} + \ketbra{\psi^\perp}{\psi^\perp} = I$.

\begin{figure}
\centering
\includegraphics[width = 65mm]{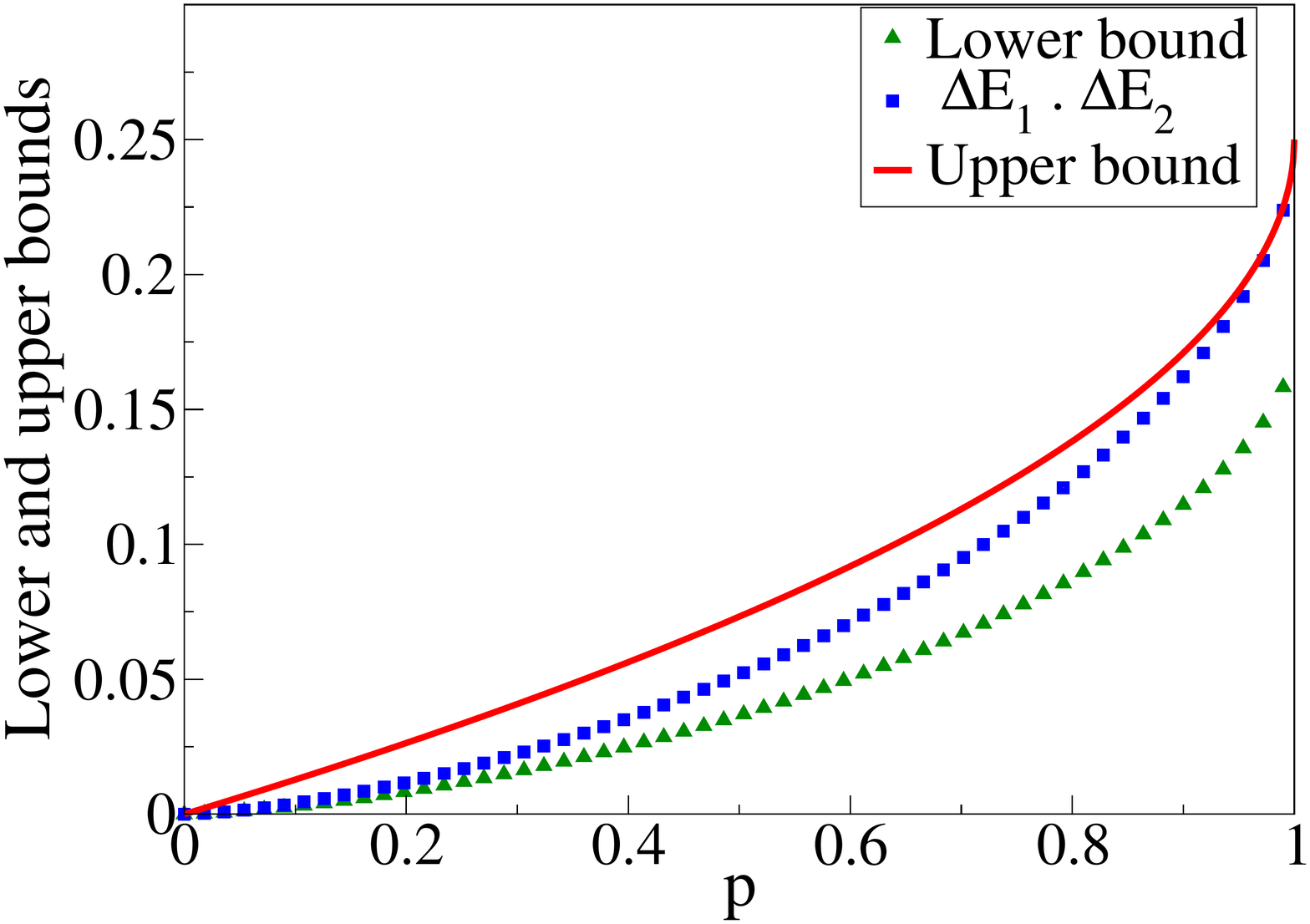}
\caption{(Color online) The lower and upper bounds on $ \Delta E_1\Delta E_2$. Here both the axes are dimensionless. In the plot blue squares show $\Delta E_1\Delta E_2$, green triangles and red solid line show the lower bound (\ref{lb}) and upper bound $\frac{1-F}{2}$, respectively on $\Delta E_1\Delta E_2$ as a function of $p$ at fixed values of $\theta = \pi/2$ and $\phi = \pi/4$.}
\label{fig2}
\end{figure}

We illustrate our uncertainty relation for Kraus operators and its relation to fidelity with the amplitude damping channel. The Kraus operators for the amplitude damping channel are given by
\begin{eqnarray}
 \label{amp}
 E_1 = \left( \begin{array}{cc}
                 1 & 0\\
		 0 & \sqrt{1-p} \end{array}\right), ~\mathrm{and}~ E_2 = \left( \begin{array}{cc}
                 0 & \sqrt{p}\\
		 0 & 0 \end{array}\right).
\end{eqnarray}
If we pass an arbitrary state $\ket{\psi} = \cos\frac{\theta}{2}\ket{0} + e^{i\phi}\sin\frac{\theta}{2}\ket{1} $ of the qubit through the amplitude damping channel, then the output state is given by
\begin{align}
 \rho &= \sum_{k=1}^{2} E_k \ketbra{\psi}{\psi}E_k^\dagger\nonumber\\
 &= \frac{1}{2} [e_1 \ketbra{0}{0} + e_2 \ketbra{1}{1} + e_3 \ketbra{0}{1} + e_3^* \ketbra{1}{0}],
\end{align}
where $e_1 = 1 + p + (1-p) \cos\theta$, $e_2 = (1-p)(1- \cos\theta)$ and $e_3 = e^{-i\phi} \sqrt{1-p} \sin\theta$. The fidelity $F= \braket{\psi}{\rho|\psi}$ is given by
\begin{align}
 F = \frac{1}{4} [ 3 + \sqrt{1-p} - p + 2 p \cos\theta  + (1-p-\sqrt{1-p})\cos2\theta].
\end{align}
Here, $S_1 = E_1$, $U_1 = I$, $S_2 = \sqrt{p}\ketbra{0}{0}$ and $U_2 = \sigma_x$. We have
\begin{align}
\label{lb}
  |_{\phi}[S_1 &\ket{\psi^\perp}\bra{\psi^\perp} S_2]^w_{\chi}|\cdot|\bra{\phi}\chi\rangle|\nonumber\\
  &= 2\cos\phi\cos^2(\frac{\theta}{2})\sin^4 (\frac{\theta}{2})\sqrt{p}[1-\sqrt{1-p}],
\end{align}
$\Delta E_1 = \cos(\frac{\theta}{2})\sin(\frac{\theta}{2})[1-\sqrt{1-p}] $ and $\Delta E_1 = \sqrt{p} \cos^2(\frac{\theta}{2})$. Fig. (\ref{fig2}) shows the bounds on $\Delta E_1\Delta E_2$ as a function of $p$ at fixed value of $\theta = \pi/2$ and $\phi = \pi/4$, which validates Eq. (\ref{final1}) for the amplitude damping channel.
\vspace{0.3 mm}

\section{Applications}
\label{apps}
In this section, we provide various interesting applications of our results. In particular, we will show a way to test the stronger uncertainty relation  \cite{Maccone2014}, to measure the product of projection operators. Interestingly, we show that our results can also be used to verify the Ramanujan's sum formula \cite{Ramanujan1915}. We, also, consider the application of our results in case of $\cal{PT}$ symmetric Hamiltonians.
\subsection{Testing stronger uncertainty relation}
Uncertainty relation plays a fundamental role in quantum mechanics and quantum information theory. Recently, the stronger uncertainty relation \cite{Maccone2014} (compared to the Robertson uncertainty relation \cite{Robertson1929}) is proved which shows that the sum of variances of two incompatible observables, $A$ and $B$ in a state $\ket{\psi}$, is given by 
\begin{align}
 \Delta A^2 + \Delta B^2  \ge \pm i\bra{\psi}[A,B]\ket{\psi} + |\braket{\psi}{A\pm i B| {\bar{\psi}} } |^2,
\end{align}
where $\ket{\bar{\psi}}$ is a state orthogonal to $\ket{\psi}$. For two canonically conjugate pair of observables such as position $X$ and momentum $P$ ($\hbar=1$ and $X, P$ are dimensionless), the stronger uncertainty relation reads as
\begin{align}
\Delta X^2 + \Delta P^2  \ge 1 + 2 |\braket{\psi}{a^{\dagger}| {\bar{\psi}} } |^2,
\end{align}
where $a^{\dagger} = (X - i P)/\sqrt{2}$ is the creation operator and it is indeed a non-Hermitian operator. The Heisenberg-Robertson uncertainty relation \cite{Robertson1929} only implies that one has $\Delta X^2 + \Delta P^2  \ge 1$, while the new relation is stronger. We will show that our protocol can be used to test this. The variances in the position and the momentum can be tested using the standard method and the last term can be actually measured using our scheme. Specifically, we will show that it is modulus squared of the weak value of positive semi-definite part of the non-Hermitian operator multiplied by a real number. First note that we can write $|\braket{\psi}{ a^{\dagger}| {\bar{\psi}} }|^2= |\braket{\bar{\psi}}{a |\psi} |^2$. Let $a= UR$, then we have $|\braket{\psi}{ a^{\dagger}| {\bar{\psi}} }|^2 =  |{_{\phi}\langle} R \rangle^w_{\psi}|^2 |\langle \phi \ket{\psi}|^2$, where $\ket{\phi}=  U^{\dagger}\ket{\bar{\psi}}$. Therefore, by measuring the non-Hermitian operator we can test the stronger uncertainty relation.

\subsection{ Measurement of product of two non-commuting projectors}
Consider measurement of $\Pi_i(B) \Pi_j(C)$, with $\Pi_i(B) = \ketbra{\psi_i}{\psi_i} $ and $\Pi_j (C) = \ketbra{\phi_j}{\phi_j} $, where $\ket{\psi_i}$ and $\ket{\phi_j}$ $(i, j = 1,2..,d)$ are eigenstates of two non-commuting Hermitian operators $B$ and $C$, respectively. The product $\Pi_i(B) \Pi_j(C)$ is a non-Hermitian operator. In fact, the average of this operator in a quantum state is nothing but 
the discrete version of the Dirac distribution \cite{Dirac1945, Chaturvedi2006, Johansen2007, Bamber2014}. We will show that our method can be applied to measure the expectation value of $\Pi_i(B) \Pi_j(C)$ using the polar decomposition and the weak measurement. For the non-Hermitian operator $A = \braket{\psi_i}{\phi_j}\ketbra{\psi_i}{\phi_j}$, let the polar decomposition be denoted by $A = UR$, where $R = |\braket{\psi_i}{\phi_j}| \ketbra{\phi_j}{\phi_j}$ and $U$ is determined by relation $ U\ket{\phi_j} = e^{i\eta} \ket{\psi_i}$, where $e^{i\eta} = \frac{\braket{\psi_i}{\phi_j}}{|\braket{\psi_i}{\phi_j}|}$. Such a unitary operator is given by
\begin{align}
 U = U(m) = e^{i\eta}\sum_{k=0}^{d-1}\ketbra{\psi_{k\oplus m}}{\phi_k},
\end{align}
with $j \oplus m = i$ and $\oplus$ denotes the addition modulo $d$. Now the expectation value of $\Pi_i(B)\Pi_j(C)$ in a state $\ket{\psi}$ is given by
\begin{align}
 \braket{\psi}{\Pi_i(B)\Pi_j(C)|\psi} = \braket{\psi}{U R|\psi}  = {_{\psi'}\langle} R \rangle^w_{\psi}\cdot \braket{\psi'}{\psi},
\end{align}
where $\ket{\psi'} = U^\dagger \ket{\psi}$. Thus, the expectation value of $\Pi_i(B)\Pi_j(C)$ in the state $\ket{\psi}$ is given by the weak value ${_{\psi'}\langle} R \rangle^w_{\psi} =\frac{\braket{\psi'}{R|\psi}} {\braket{\psi'}{\psi}}$ of $R$ multiplied by a complex number $\braket{\psi'}{\psi}$.

The weak average  \cite{Lundeen2012}, without postselection, of a non-Hermitian operator $A$ in a state $\rho$ is equal to, in general complex, expectation value of $A$ in the state $\rho$, i.e., $\langle A^w \rangle_\rho = \mathrm{Tr}[A\rho]$. Following \cite{Lundeen2005, Bamber2014}, one can devise an experimental method to measure this complex expectation value of $A$. The method is shown only for non-Hermitian operators which are product of non commuting Hermitian operators. In this method the interaction Hamiltonian for system and apparatus is designed to be $H = g\sum_{i=1}^{N}A_i \otimes P_i$ in order to measure expectation value $\langle \prod_i A_i\rangle$. Unlike our method, which can be used to measure expectation value of any non-Hermitian operator, the earlier method is applicable only to the cases of non-Hermitian operators which are product of non-commuting Hermitian operators. Next, we consider an example of non-Hermitian operator which is not product of Hermitian operators.

\subsection{Average of the creation operator and the Ramanujan sum} 
Let us consider the creation and the annihilation operators for a single mode electromagnetic field. The polar decompositions of the creation and the annihilation operators are given by Eq. (\ref{polr-decop-cre-ann}). Now consider the expectation value of the creation operator in a general state from the $(s+1)$ dimensional Hilbert space spanned by phase states, given in Eq. (\ref{phase-states}). This is given by 
\begin{align}
 \braket{\psi}{\hat{a}^\dagger|\psi} &= \braket{\psi}{\sqrt{\hat{N}} e^{-i\hat{\phi}_\theta}|\psi} = {_\psi\langle \sqrt{\hat{N}} \rangle}^w_{\chi} \cdot \braket{\psi}{\chi},
\end{align}
where $\ket{\chi} = e^{-i\hat{\phi}_\theta}\ket{\psi}$. Thus, by measuring the weak value of square root of number operator with preselection in the state $\ket{\chi}$ and postselection in the state $\ket{\psi}$ leads to the expectation value of $\hat{a}^\dagger$ in the state $\ket{\psi}$. Consider a general state $\ket{\psi} = \sum_{m = 0}^{s} c_m \ket{m}$ with $\sum_{m=0}^{s}|c_m|^2 = 1$. Here $\ket{\chi} = e^{-i\hat{\phi}_\theta}\ket{\psi} =\sum_{m=1}^{s} c_{m-1}\ket{m} + c_s e^{-i (s+1)\theta_0}\ket{0}$ and ${_\psi\langle \sqrt{\hat{N}} \rangle}^w_{\chi} = \frac{\sum_{m = 1}^{s}c_{m-1}c_m^*\sqrt{m}}{c_s c_0^* e^{-i(s+1)\theta_0} + \sum_{m = 1}^{s}c_{m-1}c_m^*}$. Therefore, we have $\braket{\psi}{a^\dagger|\psi} = \sum_{m = 1}^{s}c_{m-1}c_m^*\sqrt{m}$. For equally superposed number state, i.e., $c_m = e^{i\nu m}/\sqrt{(s+1)}$, we have
\begin{align}
\label{sqs}
 \braket{\psi}{a^\dagger|\psi} = \frac{e^{-i\nu}}{s+1} \sum_{m = 1}^{s}\sqrt{m}.
\end{align}
Using the result that for any real number $r$ with $r\geq 1$ and positive integer $n$
\begin{align}
 \sum_{m = 1}^{s}{m}^{1/r} = \frac{r}{r+1}(s+1)^{\frac{r+1}{r}} -\frac{1}{2}(s+1)^{\frac{1}{r}} - \Phi_s(r),
\end{align}
where $\Phi_s(r)$ is a function of $r$ with $s$ as a parameter and is bounded between $0$ and $1/2$ \cite{Ramanujan1915, Shekatkar2012}. Putting $r = 2$ in the above formula, we get $\sum_{m = 1}^{s}\sqrt{m} = \frac{2}{3}(s+1)^{3/2} - \frac{1}{2}(s+1)^{1/2} - \Phi_s(2)$, where $0\leq \Phi_s(2)\leq 1/2$. Therefore, we have $\braket{\psi}{a^\dagger|\psi} = \frac{e^{-i\nu}}{s+1} 
 [ \frac{2}{3}(s+1)^{3/2} - \frac{1}{2}(s+1)^{1/2} - \Phi(1/2) ]$. Interestingly, one can invert Eq. (\ref{sqs}) to get
\begin{align}
 \sum_{m = 1}^{s}\sqrt{m} = \frac{e^{i\nu}}{s+1} \braket{\psi}{a^\dagger|\psi}.
\end{align}
Therefore, one can use the expectation value of the creation operator employing our method based on weak measurements to estimate the sum of the square roots of the first $s$ natural numbers and then this result can be compared to the Ramanujan formula \cite{Ramanujan1915} for the above series. This is another interesting application of our formalism.

\subsection{Measurement of ${\cal {PT}}$ symmetric Hamiltonian}
There exists a class of Hamiltonians which are non-Hermitian and yet they possess real eigenvalues when they respect unbroken $\mathcal{PT}$ symmetry \cite{Bender1998,  Bender2002}. However, in general they posses non-normalizable eigenstates and complex eigenvalues, so one may think that we cannot measure their expectation values. But using our formalism which is based on weak measurements one can in principle measure them. Therefore, given a non-Hermitian Hamiltonian, one can check whether its expectation value indeed gives a complex number.

The simplest example of a general  $\mathcal{PT}$ symmetric Hamiltonian in $2 \times 2$, is given by \cite{Bender2002}
\begin{eqnarray}
 H = \left( \begin{array}{cc}
                 r e^{i \theta} & t\\
		 s & r e^{-i \theta} \end{array}\right),
\end{eqnarray}
where $r$, $s$, $t$, $\theta$ are the real parameters. The eigenvalues are given by $\epsilon_{\pm} = r \cos \theta \pm \sqrt{st -r^2 \sin^2 \theta}$ and corresponding eigenstates of this Hamiltonian are given by 
\begin{eqnarray}
\ket{\epsilon_+} = \frac{1}{\sqrt{ 2 \cos \alpha}} 
	\left( \begin{array}{r} e^{i\alpha/2} \\
	e^{-i\alpha/2} \end{array} \right);~~ 
\ket{\epsilon_-} = \frac{1}{\sqrt{ 2 \cos \alpha}} 
	\left( \begin{array}{r} e^{-i\alpha/2} \\
	-e^{i\alpha/2} \end{array} \right), \nonumber
\end{eqnarray}
where $\alpha$ is defined by the relation $\sin \alpha = \frac{r}{\sqrt{s t}} \sin \theta$. Let the polar decomposition of $H$ be $H = U R$, where $R = \sqrt{H^\dagger H}$ and for $r^2 \neq s t$,
$U = H R^{-1}$. Here,
\begin{eqnarray}
R^2 = \left( \begin{array}{cc} 
 	r^2 + s^2 & r(s+t)e^{-i\theta} \\
	r(s+t)e^{i\theta} & r^2 + t^2 \end{array} \right).
\end{eqnarray}
Therefore, the positive semi-definite operator $R$ for the non-Hermitian operator $H$ is given by
\begin{eqnarray}
R = \frac{1}{2\sqrt{2}A}\left( \begin{array}{cc} 
 	R_{11} & R_{21} \\
	R_{12} & R_{22} \end{array} \right),
\end{eqnarray}
where $R_{11} = (A - (s-t))B_- + (A + (s-t))B_+ $, $R_{12} = 2r(B_+ - B_-)e^{-i\theta} $, $R_{21} = 2r(B_+ - B_-)e^{i\theta}$ and $R_{22} = (A + (s-t))B_- + (A - (s-t))B_+ $ with
\begin{align}
 &A = \sqrt{ 4 r^2 + (s-t)^2}\\
 &B_{\pm} = \sqrt{2r^2 + s^2 + t^2 \pm (s+t)A}.
\end{align}
Now, we have
\begin{eqnarray}
R^{-1} = \frac{1}{\sqrt{2}A B_+ B_-} \left( \begin{array}{cc} 
 	S_{11} & S_{12} \\
	S_{21} & S_{22} \end{array} \right),
\end{eqnarray}
where $S_{11} = (A + (s-t))B_- + (A - (s-t))B_+ $, $S_{12} = -2r(B_+ - B_-)e^{-i\theta}$, $S_{21} = -2r(B_+ - B_-)e^{i\theta} $, and $S_{22} = (A - (s-t))B_- + (A + (s-t))B_+ $. Using $U = H R^{-1}$, we have
\begin{eqnarray}
U = \frac{1}{\sqrt{2}A B_+ B_-} \left( \begin{array}{cc} 
 	U_{11} & U_{12} \\
	U_{21} & U_{22} \end{array} \right),
\end{eqnarray}
where $U_{11} = [(A + (s-t))B_- + (A - (s-t))B_+ - 2t(B_+ - B_-)]re^{i\theta}$, $U_{12} = t[(A - (s-t))B_- + (A + (s-t))B_+ - 2r^2(B_+ - B_-)$, $U_{21} = s[(A + (s-t))B_- + (A - (s-t))B_+ - 2r^2(B_+ - B_-)$ and $U_{22} = [(A - (s-t))B_- + (A + (s-t))B_+ - 2s(B_+ - B_-)]re^{-i\theta} $. For the special case of $s = t$, $r\neq \pm s$ and $r > s$, we have $H = U R$, where
\begin{eqnarray}
R = \left( \begin{array}{cc} 
 	r & se^{-i\theta}  \\
	se^{i\theta} & r \end{array} \right),~~U = \left( \begin{array}{cc} 
 	e^{i\theta} & 0  \\
	0 & e^{-i\theta} \end{array} \right).
\end{eqnarray}
For other special case of $s = t$, $r\neq \pm s$ and $r < s$, we have $H = U R$, where
\begin{eqnarray}
R = \left( \begin{array}{cc} 
 	s & re^{-i\theta}  \\
	re^{i\theta} & s \end{array} \right),~~U = \left( \begin{array}{cc} 
 	0 & 1  \\
	1 & 0\end{array} \right).
\end{eqnarray}
Now the expectation value of $A$ in a general single qubit state $\ket{\psi} = \cos(\eta/2) \ket{0} + e ^{i\xi} \sin(\eta/2) \ket{1}$ is given by
\begin{align}
 \braket{\psi}{H|\psi} = r\cos\theta + s \cos \xi \sin\eta + i r \sin\theta \cos\eta.
\end{align}
The above expectation value of $A$ in the state $\ket{\psi}$ can be realized if we experimentally measure the Hermitian operator $R$ with the preselection in the state $\ket{\psi}$ and postselection in the state $U^\dagger \ket{\psi}$ and then multiply the weak value thus obtained with the complex number $\braket{\psi}{U|\psi}$.

\vspace{0.5 cm}
\section{Conclusion}
\label{conclusion}
In this paper, we have addressed the question of experimental feasibility of measuring the expectation value of {\em any non-Hermitian operator} in a pure quantum state.
We show that the expectation value of a non-Hermitian operator in a quantum state is equal to the weak value of the positive semi-definite part of the operator, modulo a complex number. Our method to measure the expectation value require a priori knowledge of both the operator to be measured and the state in which it is being measured. However, this situation may arise naturally in many contexts, therefore our method is widely applicable. We have provided several examples to illustrate this technique. In particular, we have provided a relation between the average of the Kraus elements of a channel and the channel fidelity. We have also applied our method to measure the expectation value of creation operator in a general state. This leads to an interesting link between the sum of the square roots of first $s$ natural numbers and the expectation value of the creation operator. Furthermore, we have proved an uncertainty relation for any two non-Hermitian operators. Our method also helps to test the stronger uncertainty relation, experimentally.
Our paper may open up the possibility of considering the non-Hermitian operators not only as a mathematical tool but also an experimental arsenal such as in scattering experiments in quantum physics and quantum information theory, in general.

\begin{center}
 {\bf ACKNOWLEDGEMENTS}
\end{center}
Uttam Singh acknowledges the research fellowship of Department of Atomic Energy, Government of India.

\bibliographystyle{apsrev4-1}
\bibliography{nh-weak-lit}

\end{document}